\documentclass[aps,prd,twocolumn,groupedaddress,showpacs,preprintnumbers,draft]
{revtex4}
\usepackage{epsf} 
\begin{document}
\preprint{BROWN-HET-1405}
\title{Back-Reaction of Cosmological Perturbations in the Infinite Wavelength
Approximation}
\author{Robert H. Brandenberger}%
	\email[Email:]{rhb@het.brown.edu}
\affiliation{Department of Physics, Brown University, Providence, RI 02912, 
USA, and\\
Perimeter Institute for Theoretical Physics, Waterloo, N2J 2W9, CANADA, and\\
Physics Department, McGill University, 3600 University
St., Montr\'eal QC, H3A 2T8, CANADA} 

\author{C.S. Lam}%
	\email[Email:]{lam@hep.physics.mcgill.ca}
\affiliation{Physics Department, McGill University, 3600 University
St., Montr\'eal QC, H3A 2T8, CANADA}

\begin{abstract}
Cosmological perturbations in an expanding universe back-react on the
space-time in which they propagate. Calculations to lowest 
non-vanishing order in perturbation theory indicate that super-Hubble-scale
fluctuations act as a negative and time-dependent cosmological constant
and may thus lead to a dynamical relaxation mechanism for the
cosmological constant. Here we present a simple model of how to
understand this effect from the perspective of homogeneous and
isotropic cosmology. Our analysis, however, also shows that an
effective spatial curvature is induced, indicating potential problems
in realizing the dynamical relaxation of the cosmological constant
by means of back-reaction.
\end{abstract}
\pacs{98.80.Cq.}
\maketitle
\def\Box{\nabla^2}  
\def\ie{{\em i.e.\/}}  
\def\eg{{\em e.g.\/}}  
\def\etc{{\em etc.\/}}  
\def\etal{{\em et al.\/}}  
\def\S{{\mathcal S}}  
\def\I{{\mathcal I}}  
\def\mL{{\mathcal L}}  
\def\H{{\mathcal H}}  
\def\M{{\mathcal M}}  
\def\N{{\mathcal N}} 
\def\O{{\mathcal O}} 
\def\cP{{\mathcal P}} 
\def\R{{\mathcal R}}  
\def\K{{\mathcal K}}  
\def\W{{\mathcal W}} 
\def\mM{{\mathcal M}} 
\def\mJ{{\mathcal J}} 
\def\mP{{\mathbf P}} 
\def\mT{{\mathbf T}} 
\def\mR{{\mathbf R}}
\def\mS{{\mathbf S}}
\def\mX{{\mathbf X}}
\def\mZ{{\mathbf Z}}
\def\eff{{\mathrm{eff}}}  
\def\Newton{{\mathrm{Newton}}}  
\def\bulk{{\mathrm{bulk}}}  
\def\brane{{\mathrm{brane}}}  
\def\matter{{\mathrm{matter}}}  
\def\tr{{\mathrm{tr}}}  
\def\normal{{\mathrm{normal}}}  
\def\implies{\Rightarrow}  
\def\half{{1\over2}}  
\newcommand{\da}{\dot{a}}
\newcommand{\db}{\dot{b}}
\newcommand{\dn}{\dot{n}}
\newcommand{\dda}{\ddot{a}}
\newcommand{\ddb}{\ddot{b}}
\newcommand{\ddn}{\ddot{n}}
\newcommand{\ba}{\begin{array}}
\newcommand{\ea}{\end{array}}
\def\be{\begin{equation}}
\def\ee{\end{equation}}
\def\bea{\begin{eqnarray}}
\def\eea{\end{eqnarray}}
\def\bs{\begin{subequations}}
\def\es{\end{subequations}}
\def\g{\gamma}
\def\G{\Gamma}
\def\vp{\varphi}
\def\mpl{M_{\rm P}}
\def\ms{M_{\rm s}}
\def\ls{\ell_{\rm s}}
\def\lp{\ell_{\rm pl}}
\def\l{\lambda}
\def\gs{g_{\rm s}}
\def\d{\partial}
\def\co{{\cal O}}
\def\sp{\;\;\;,\;\;\;}
\def\spa{\;\;\;}
\def\r{\rho}
\def\dr{\dot r}
\def\dt{\dot\varphi}
\def\e{\epsilon}
\def\k{\kappa}
\def\m{\mu}
\def\n{\nu}
\def\om{\omega}
\def\tn{\tilde \nu}
\def\p{\phi}
\def\vp{\varphi}
\def\P{\Phi}
\def\r{\rho}
\def\s{\sigma}
\def\t{\tau}
\def\x{\chi}
\def\z{\zeta}
\def\a{\alpha}
\def\b{\beta}
\def\de{\delta}
\def\cth{c_{\theta}}
\def\sth{s_{\theta}}
\def\bra#1{\left\langle #1\right|}
\def\ket#1{\left| #1\right\rangle}
\newcommand{\stt}{\small\tt}
\renewcommand{\theequation}{\arabic{section}.\arabic{equation}}
\newcommand{\eq}[1]{equation~(\ref{#1})}
\newcommand{\eqs}[2]{equations~(\ref{#1}) and~(\ref{#2})}
\newcommand{\eqto}[2]{equations~(\ref{#1}) to~(\ref{#2})}
\newcommand{\fig}[1]{Fig.~(\ref{#1})}
\newcommand{\figs}[2]{Figs.~(\ref{#1}) and~(\ref{#2})}
\newcommand{\GeV}{\mbox{GeV}}
\def\ricci{R_{\m\n} R^{\m\n}}
\def\riemann{R_{\m\n\l\s} R^{\m\n\l\s}}
\def\triemann{\tilde R_{\m\n\l\s} \tilde R^{\m\n\l\s}}
\def\tricci{\tilde R_{\m\n} \tilde R^{\m\n}}

\section{Introduction}

The cosmological constant problem is one of the most important 
problems of theoretical physics. The issue is to find an
explanation for the fact that the value of the cosmological constant 
in the present universe - inferred from observational upper bounds on
the contribution of the cosmological constant to the ``energy''
content of the present universe - is much 
smaller than what can be deduced from theoretical estimates for
the vacuum energy density $\rho_{vac}$. From considerations of quantum field
theory one would expect the vacuum energy density to be of the
order $M^4$, where $M$ is the ultraviolet cutoff of the theory. In 
non-supersymmetric theories, the mismatch is by a factor of
$10^{-120}$ (see e.g. \cite{Weinberg,Press} for reviews). In
supersymmetric models, $\rho_{vac}$ is of the order $M_s^4$,
where $M_s$ is the scale of supersymmetry breaking. Thus, the
theoretical predictions still exceeds the observational bounds
by about $60$ orders of magnitude.

Recent data, both from supernovae observations \cite{SNproject}
and also from cosmic microwave anisotropy
measurements (see \cite{WMAP} for the most recent observational
results) indicate that the universe is right now entering a new
stage of acceleration, indicating the presence of something
that acts as an effective cosmological constant with the value 
$\Lambda_{eff} \sim 10^{-120} M_{p}^4$. Thus, there now appear
to be two aspects of the cosmological constant problem, firstly
why the cosmological constant is so tiny compared to theoretical
expectations (the ``old cosmological constant'' problem - using
Weinberg's language \cite{Weinberg2}) and the problem
of why, given that it is so small, the cosmological constant 
does not exactly vanish, but is becoming visible
precisely at the present time of cosmic history (the ``new'
cosmological constant'' or ``coincidence'' problem) .

Independent studies of the back-reaction effects of long-wavelength
gravitational waves \cite{WT} and of long-wavelength cosmological
perturbations \cite{ABM} on the background geometry of space-time
have led to a scenario \cite{WT2,RHBrev} in which these fluctuations
lead to a dynamical relaxation of an initially large bare
cosmological constant. The large initial cosmological constant
leads to a period of primordial inflation. During this period,
quantum vacuum fluctuations of both gravitons and scalar metric
fluctuations are stretched beyond the Hubble radius, thus
generating a large phase space of long-wavelength modes (see 
e.g. \cite{MFB,RHBrev2} for reviews of the theory of cosmological
fluctuations). In an inflationary background, it 
can be shown that these modes have a back-reaction
effect on the local geometry which is analogous to that of a negative
cosmological constant (this effect may not be physically measurable
in models with only one matter field \cite{Unruh,AW,GB1} - see, however,
\cite{Fabio} for a different conclusion -, but it is definitely
physically measurable in models with two or more matter fields
\cite{GB2}). This opens up the possibility that back-reaction
may lead to a dynamical relaxation of the bare cosmological constant.

In this note we wish to focus on a major problem with the past analyses 
of the gravitational back-reaction mechanism (see e.g. \cite{Unruh} for
a discussion of these and other problems). Namely, if the back-reaction
is due to extremely long wavelength modes (for which contributions
from spatial gradient terms are negligible), then it should be possible
to understand the physical effect from the point of view of the
local equations of homogeneous and isotropic cosmology \footnote{We
thank Alan Guth and Alex Vilenkin for stressing this point to us.}.

We present a simple way of modeling the back-reaction effects
of long wavelength fluctuations in terms of the equations of
homogeneous and isotropic cosmology
(see also \cite{WT3} for an earlier work on the modeling of
non-perturbative effects). Our result is that the back-reaction of
long wavelength cosmological fluctuations in an inflationary
background appears as a local
fluctuation of the cosmological constant. On average, the
effect leads to a {\it decrease} of the cosmological constant. 
However, in a general background, a positive correction to the
spatial curvature is induced. This may lead to stringent
constraints on the back-reaction mechanism.

In inflationary cosmology the phase space of super-Hubble
fluctuations builds up over time since the wavelengths of
modes with fixed comoving wavelength red-shift relative to the
Hubble radius. We incorporate this result into our local
cosmological model by taking the density perturbation to increase
over time, which is modeled by a decreasing cosmological constant.
As discussed in the perturbative framework in \cite{RHBrev},
this leads to a {\it dynamical scaling solution} in which
\begin{equation} \label{fixed}
\Omega_{\Lambda_{eff}}(t) \sim 1 \, ,
\end{equation}
(where $\Omega_X = \rho_x / \rho_c$ is a measure of the contribution
of $X$ to the closure density $\rho_c$ of the universe, $\rho_X$
denoting the effective energy density in some $X$ ``matter'')
at all sufficiently
late times $t$. Thus, this dynamical relaxation mechanism would 
automatically address both the old and the new cosmological constant
problems.

The reason why one hopes to obtain the dynamical fixed point 
solution (\ref{fixed}) is as follows: as the phase
of inflation proceeds, the phase space of infrared modes 
builds up. Since long-wavelength fluctuations are frozen,
the contribution of an individual mode to the effective
cosmological constant does not decrease in absolute magnitude.
Thus, the total back-reaction effect builds up gradually,
and the effective cosmological constant, which is the sum
of the positive bare cosmological constant and the induced
negative back-reaction term, decreases. However, the sum cannot
drop to zero, since before this happens the energy density
$\rho_{\Lambda}$ corresponding to the {\it effective} cosmological
constant will drop below the matter energy density
\footnote{See \cite{Anupam} for a recent study of the
graceful exit from inflation in this scenario which provides
naturally a large matter energy density at late times. Note also that
in the absence of pre-inflationary matter, the relaxation
process of $\Lambda$ would evolve into a period of deflation
\cite{TW03}, from which the exit into an expanding universe would be more
involved.}. As soon
as this happens, inflation will end, the phase space of infrared
modes will cease to increase, and the back-reaction contribution
to $|\Lambda_{eff}|$ levels off. Since the universe is still
expanding, the matter energy density $\rho_{m}$ (where ``matter''
here stands for both cold matter and radiation) continues to
decrease, thus allowing $\rho_{\Lambda}$ to start dominating
again, enabling the back-reaction effect to again increase
in strength. Thus, as explained in detail in \cite{RHBrev}, we
expect that at all sufficiently late times $\rho_{\Lambda} / \rho_m$
will be oscillating in time about the value $0.5$ \footnote{Note
that a similar scenario where the effective cosmological constant
fluctuates and whose associated energy density always tracks
that of matter emerges from the causal set approach to quantum
gravity \cite{Sorkin,SorDod}. Whereas in our approach the
effective cosmological constant must remain positive, in the
causal set approach its sign fluctuates.}.

\vskip0.5cm
\section{Back-Reaction of Cosmological Perturbations: A Brief Review}

Both gravitational waves and cosmological fluctuations (scalar
metric fluctuations) carry energy and momentum and thus
affect the background in which they propagate. One
way to describe this back-reaction effect 
(see \cite{Brill} for the initial work on the back-reaction of
gravitational waves) is by defining
an energy-momentum tensor which describes the effects of the
fluctuations on the background geometry, to leading order in perturbation 
theory (the expansion parameter being the amplitude of the metric
fluctuations measured in momentum space).

In the case of cosmological perturbations,
the back-reaction formalism was initially developed in
\cite{ABM}. Small fluctuations of the
metric and the matter fields about a classical homogeneous and isotropic 
background are introduced. The metric and matter fields including
these linear perturbations (which are taken to satisfy the linear perturbation
equations) are inserted into the Einstein equations. These
equations are then expanded to second order. The 
Einstein equations are not satisfied at second
order, and it is necessary to add back-reaction terms which are
quadratic in $\epsilon$ (the relative amplitude of the linear
fluctuations). In particular, one needs to add a
second order correction to the zero mode of the metric. The
sum of the background metric plus the quadratic zero mode
correction defines a new homogeneous metric $g_{\mu\nu}^{(0,br)}$
which takes into account the effects of back-reaction to this
order. This new metric obeys the modified equations
\begin{equation}
G_{\mu\nu}(g_{\alpha\beta}^{(0,br)}) = 3 \kappa \left[T_{\mu\nu}^{(0)}
+\left\langle T^{(2)}_{\mu\nu} - {1 \over {8 \pi G}}G^{(2)}_{\mu\nu}\right
\rangle\right]\,,
\end{equation}
where $\kappa = 8 \pi G / 3$, the constant $G$ denoting Newton's
gravitational constant, $G_{\mu \nu}$ is the Einstein tensor, 
$T_{\mu \nu}$ is the
energy-momentum tensor of matter, the pointed brackets stand for spatial 
averaging \footnote{It was shown in \cite{ABM} that the
back-reaction equation is covariant under linear space-time
coordinate transformations.}, 
and the superscripts indicate the order in perturbation theory.
The terms inside the pointed brackets form the
{\it effective energy-momentum tensor} $\tau_{\mu \nu}$ of cosmological
perturbations. Note that all
Fourier modes of the fluctuating fields contribute to
the effective energy-momentum tensor for back-reaction: the
effect of perturbations is cumulative. Thus, back-reaction can
be important even if the relative magnitude of each metric
fluctuation mode is small (as observations indicate) as long as
there is a sufficiently large phase space of infrared modes, i.e.
as long as the period of primordial inflation is sufficiently long.

As was found in \cite{ABM}, in a background space-time corresponding
to slow-roll inflation, the terms which dominate the effective energy-momentum
tensor $\tau_{\mu \nu}$ of super-Hubble modes takes the 
form of a {\it negative cosmological constant} (see \cite{Ghazal} for
a recent discussion of next to leading terms)
\begin{equation}
\label{result}
p_{br}=-\rho _{br} \,\,\, {\rm with} \,\,\, \rho_{br} < 0 \, ,
\end{equation}
where $\rho_{br}$ and $p_{br}$ stand for the energy density and
pressure associated with $\tau_{\mu \nu}$. In hindsight, 
it is easy to understand why the back-reaction of
infrared modes of cosmological perturbations acts as a negative
cosmological constant. For long wavelength fluctuations, all terms
in $\tau_{\mu \nu}$ involving spatial derivatives
are negligible. Since on long wavelengths the amplitude of the
metric fluctuations is frozen (see e.g. \cite{MFB}), and assuming
that the background matter fields are slowly rolling,
all terms involving temporal derivatives of the fluctuation variables
are also negligible. The only terms which survive are gravitational potential
energy terms, terms which act as a cosmological constant. 
Since matter fluctuations produce negative potential
wells, and since for infrared modes the negative gravitational energy
dominates over the positive matter energy, the total effective energy
density is negative. 

The crucial observation is that the absolute value of $\rho_{br}$ 
increases as a function of time. This is because, in an inflationary 
universe, due to the accelerated expansion of space, the phase space
of infrared modes increases in time as wavelengths exit the 
Hubble radius. 
 
Since it is the total metric and not the background metric only
which determines observables, it was suggested \cite{RHBrev} 
on the basis of the above results that the effective 
cosmological constant at time $t$ is given by
\begin{equation}
\label{result1}
\Lambda_{eff}(t)=\Lambda_{0}+\rho_{br}\, ,
\end{equation}
where the second term on the right hand side is negative and has
an absolute value which is increasing in time. Since mode by
mode the magnitude of the square of the fluctuation amplitude is tiny, 
the back-reaction contribution to the effective cosmological 
constant is very small compared to 
the bare cosmological constant $\Lambda_{0}$ during the early
stages of inflation. However, if inflation lasts long enough (as
will be the case in many inflationary models in the class of
``chaotic'' inflation \cite{chaotic}), then, before the
homogeneous scalar field $\varphi$ has ended the slow-rolling phase, the
back-reaction contribution to the effective cosmological constant 
will cancel the initial bare value $\Lambda_{0}$. 
However, as described in the Introduction, the energy
density associated with $\Lambda_{eff}(t)$ can in fact never
become negative, and at late times one will obtain a dynamical
scaling solution in which the energy density associate with
$\Lambda_{eff}(t)$ tracks the matter energy density.

There are key conceptual and technical issues to be resolved
before the above speculations can be considered to be well-founded.
The problem we will focus on in the next section is the
following: If the back-reaction effect is due to 
very long wavelength fluctuations, and the effect is to be
physically measurable by a local observer, then it should be
possible to understand the physics making use of only the local
cosmological equations \footnote{See \cite{Niayesh} for a previous
approach to this problem.}.

\vskip0.5cm
\section{Local Model of Back-Reaction}

We now propose a way of understanding the origin of back-reaction
using the local equations of homogeneous and isotropic cosmology.
We start with the metric including linear cosmological perturbations
which in longitudinal gauge and for matter without anisotropic stress
takes the form (in the absence of spatial curvature) \cite{MFB} 
\be
ds^2 \, = \, (1 + 2 \phi) dt^2 - a^2(t)(1 - 2 \phi) d{\bf x}^2 \, ,
\ee
where $\phi({\bf x}, t)$ denotes the metric fluctuation, $a(t)$ is the
scale factor of the background cosmology, $t$ is the background time
coordinate, and ${\bf x}$ are the comoving spatial coordinates. Note
that for fluctuations with wavelengths larger than the Hubble radius,
the dominant mode of $\phi$ is independent of time if the equation of
state of the background is constant \cite{MFB}.

As a first step towards applying the equations of homogeneous
and isotropic cosmology to a local patch of the universe, we define
a new time coordinate ${\tilde t}$ via
\be \label{newtime}
d{\tilde t}^2 \, = \, (1 + 2 \phi) dt^2 \, .
\ee
The first local cosmological equation determines the Hubble expansion rate
and reads
\be \label{FRW1}
({{da / d{\tilde t}} \over a})^2 \, = \, \kappa {\tilde \rho} - 
{k \over {a^2}} + {{\Lambda} \over 3} \, ,
\ee
where ${\tilde \rho} = \rho + \delta \rho$ is the local energy density, and
$k$ is the curvature constant ($k = 1$, $k = 0$ and $k = -1$ denoting the 
cases of a closed, spatially flat and open universe, respectively). 
The second local equation gives the acceleration
\be \label{FRW2}
{{d^2a / d{\tilde t}^2} \over a} \, = \, 
- {{\kappa} \over 2} ({\tilde \rho} + 3{\tilde p})
+ {{\Lambda} \over 3} \, ,
\ee
where ${\tilde p} = p + {\delta p}$ stands for the pressure density.

We wish to apply these equations in an Inflationary universe. Since
inflation typically renders the universe spatially flat, we will set
the background value of $k$ to zero. Let us
ask how one can model the effects of a matter density perturbation
$\delta \rho$ and its associated metric fluctuations in terms of the 
above local equations. We wish to
model fluctuation modes with wavelength far larger than the 
Hubble radius. On these scales, the dominant mode of the
metric fluctuation variable $\phi$ is time-independent.
 
Inserting (\ref{newtime}) into (\ref{FRW2}), 
remembering that $\phi$ is time-independent,
and expanding to second order in $\phi$, we obtain
\be
\bigl( 1 - 2 \phi + 4 \phi^2 \bigr) {{\ddot a} \over a} \, = \,
- {{\kappa} \over 2} \bigl( \rho + 3 p \delta \rho + 3 \delta p \bigr) 
+ {1 \over 3} \Lambda \, .
\ee
The terms linear in $\phi$ and $\delta \rho$ vanish upon spatial averaging,
leaving us (after making use of the background equation) 
with the result 
\be
{{\ddot a} \over a} \, = \,
\kappa \rho  + {1 \over 3} \Lambda - 4 <\phi^2> \bigl( {{\ddot a} \over a} \bigr)_0
\ee
(where the pointed brackets standing for averaging and the subscript $0$ indicates
that the corresponding quantity is to be evaluated using the background metric)
which implies that a long wavelength cosmological perturbation
acts, when measured in terms of the background time coordinate $t$, 
like a local correction to the cosmological constant whose value is
\be \label{adjust1}
\delta \Lambda \, = \, - 12 \bigl( {{\ddot a} \over a}(t) \bigr)_0 <\phi^2(t)> \, ,
\ee
i.e. to a reduction in the locally measured value of the cosmological
constant. Note that this result does not, in fact, depend on the
equation of state of the background which we have assumed.

If we insert (\ref{newtime})
into the other equation of motion (namely (\ref{FRW1})), expand
to second order in $\phi$, and make use of (\ref{adjust1}), then we obtain
\be \label{adjust2}
\bigl({{\dot a} \over a}\bigr)^2 \, = \, \kappa \rho + 
{1 \over 3} \bigl( \Lambda + \delta \Lambda \bigr) - 
4 <\phi^2> \bigl[ ({{\dot a} \over a})^2 - {{\ddot a} \over a} \bigr]_0 \, . 
\ee
The final term on the right hand side of (\ref{adjust2}) acts like a 
local increase in the spatial curvature. 
During the phase of inflation, this term vanishes, but it no longer vanishes
after inflationary reheating.

As mentioned in the Introduction, in inflationary cosmology 
the accelerated expansion of space leads to a continuous growth
of the phase space of infrared modes. To model the local effects of
the infrared modes in (\ref{adjust1}), we use 
\be \label{posspace}
\phi^2(t, {\bf x}) \, = \, \int d^3{\bf k} |\phi(t, {\bf k})|^2 \, ,
\ee
where $\phi(t {\bf k})$ is the amplitude of the fluctuation mode
with wavenumber ${\bf k}$, and the integral runs over all wavelengths
larger than the Hubble radius. In slow roll inflation,
the spectrum of fluctuations is nearly scale-invariant \cite{Mukh},
\be
k^3 |\phi(k)|^2 \, \simeq \, {\rm const} \, .
\ee
Thus, the integral in (\ref{posspace}) over the infrared modes
depends logarithmically on $k_{max}(t)$, the comoving wavenumber
corresponding to the Hubble radius. In turn, in inflationary cosmology
$k_{max}$ is increasing exponentially, and thus the integral
grown linearly in time. Thus, the locally measured cosmological constant 
will continue to decrease.

Eventually, the energy density associated with the effective cosmological 
constant will drop below the energy density of the matter/radiation
fluid. At this time, inflation will terminate. The phase
space of infrared modes begins to decrease again, leading to a
slow decrease of the integral in (\ref{posspace}). More importantly,
it follows from (\ref{adjust1})) that $|\delta \Lambda|$ decreases in
time since $H(t)$ is decreasing.
Thus, the cosmological
constant will increase again until its associated energy density
becomes larger than the matter energy density, after which the inflationary
dynamics takes over again.

Combining the arguments in the two previous paragraphs, we see
that our local analysis allows us to understand the origin
of the dynamical fixed point (\ref{fixed}). However, it also
leads to a potential problem of the attempt to use back-reaction
to relax the cosmological constant: it appears that back-reaction
also generates a local spatial curvature term.
  
\vskip0.5cm

\section{Discussion and Conclusions} 

In this Note we have demonstrated that the back-reaction of
long wavelength (super-Hubble) cosmological perturbations on
the locally measured background geometry can be understood in
the quasi-homogeneous approximation (i.e. neglecting all spatial
gradients) using the equations of homogeneous and isotropic
cosmology alone. 

When applied to inflationary cosmology, we have recovered the results 
of \cite{ABM}, namely that super-Hubble fluctuation modes
affect the background space-time like an additional matter source
with an equation of state of a negative cosmological constant whose
absolute value is increasing in time because the phase space of
infrared modes is increasing. This supports the conjectures of
\cite{RHBrev} that back-reaction of infrared modes leads to a
dynamical relaxation of any bare cosmological constant. Note that
the continued quantum generation of ultraviolet fluctuations is
key to the relaxation. Taking only initial classical fluctuations
into account, one would simply obtain a small renormalization of the
cosmological constant (see also \cite{Mottola,Traschen} for related work).

However, from (\ref{adjust2}) we see that in a general background
cosmology the back-reaction terms also induce a term which acts
like a positive correction to the spatial curvature. This term
vanishes during a phase of quasi-exponential inflation, but it
does not vanish in late time cosmology. Since there are stringent
observational constraints on the spatial curvature of the present
universe, our mechanism may already be tightly constrained. 
One must keep in mind, however, that the observational constraints
come from the properties of the cosmological fluctuations. Thus, before
drawing definite conclusions, it is necessary to study how
back-reaction effects the evolution of fluctuations.

A key issue is whether the back-reaction effect calculated here
(and in the previous work of \cite{ABM})
using the background time coordinate $t$ can be physically measured
(i.e. identified as a back-reaction effect by some physical clock).
This issue was raised most forcefully in \cite{Unruh}. To answer
this question, one has to calculate the back-reaction on a physical
observable and express the result in terms of a physical clock such
as $\varphi$. Work of \cite{Niayesh,AW,GB1} has shown that the
effect of the leading infrared terms in the effective energy-momentum
tensor are in fact not physically measurable (see, however, \cite{Fabio}
for a differing view). However, they are physically measurable in
models with an extra physical clock, e.g. a second scalar matter
field $\chi$ \cite{GB2}. These conclusions also apply to our present
work. 

\vskip0.5cm
\noindent{\it Acknowledgments:}  One of us (RB) is
grateful to Alan Guth for extensive discussions emphasizing
the need for a local analysis of back-reaction. We also
thank Niayesh Afshordi and Richard Woodard for comments on
the manuscript. RB is supported in 
part by the US Department of Energy under Contract DE-FG02-91ER40688, 
TASK~A. He thanks the Perimeter Institute for their gracious hospitality 
and financial support during the course of the work on
this project. The work of CSL is supported by NSERC
(Canada) and by the Fonds de Recherche sur la Nature et les
Technologies du Qu\'ebec.

\end{document}